# ChemiQ: A Chemistry Simulator for Quantum Computer

Qingchun Wang,[1] Huan-Yu Liu,[2,3,4] Qing-Song Li,[2,3,4] Jianyu Zhao,[5] Deping Huang,[5] Qingmin Man,[5] Qiankun Gong,[5] Ye Li,[5] Menghan Dou,[5] Yu-Chun Wu,[1,2,3,4] Guo-Ping Guo[1,2,3,4,5] *

[1]Institute of Artificial Intelligence, Hefei Comprehensive National Science Center, Hefei, Anhui 230088, China

[2]CAS Key Laboratory of Quantum Information, University of Science and Technology of China, Hefei, Anhui, 230026, People's Republic of China

[3]CAS Center For Excellence in Quantum Information and Quantum Physics, University of Science and Technology of China, Hefei, Anhui, 230026, People's Republic of China

[4]Hefei National Laboratory, University of Science and Technology of China, Hefei 230088, China

[5]Origin Quantum Computing Company Limited, Hefei, Anhui 230026, People's Republic of China

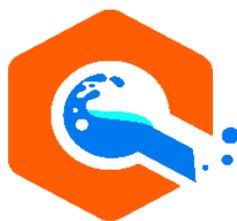

**ABSTRACT:** Quantum computing, an innovative computing system carrying prominent processing rate, is meant to be the solution to problems in many fields. Among these realms, the most intuitive application is to help chemical researchers correctly describe strong correlation and complex systems, which are the great challenge in current chemistry simulation. In this paper, we will present a standalone quantum simulation tool for chemistry, ChemiQ, which is designed to assist people in carrying out chemical research or molecular calculation on real or virtual quantum computers. Under the idea of modular programming in C++ language, the software is designed as a full-stack tool without third-party physics or chemistry application packages. It provides services as follows: visually construct molecular structure, quickly simulate ground-state energy, scan molecular potential energy curve by distance or angle, study chemical reaction, and return calculation results graphically after analysis.

**Keywords**: quantum computing, chemistry simulation, quantum chemistry, variational quantum eigensolver, unitary coupled cluster

## I. Introduction

Electronic computer has brought great convenience to people's life. Current manufacture has developed nano-scaled technique to support complicated computation, which is widely spread in today's academic researches and for commercial purposes. However, according to Moore's law[1] and theories in quantum mechanics, the number of silicon transistors within a chip would eventually converge to its limit and therefore approach to quantum effect, which especially impacts the fields of quantum chemistry, geographic remote sensing, weather prediction, and so on. Differing from classical computers based on binary computation, quantum computation[2] based on quantum mechanics has an exponential acceleration by using the superposition or entanglement of quantum bits (qubits).[3-6] Consequently, it has attracted extensive attention in recent years.[7-11]

Due to the developing technique in manufacturing practical quantum computers, the number of qubits we can utilize in the noisy intermediate-scale quantum (NISQ) era[12] is limited, so it is very helpful for us to search on the application of quantum simulation. First, we can learn and analyze the performance of quantum systems quickly and economically due to the simplicity and controllability of simulated computation. Secondly, we can develop some new algorithms or computational frameworks for effective simulation. Further, we can build the whole quantum software ecosystem. Finally, in turn, engineers are guided to design and manufacture quantum computers with high performance.

Chemistry simulation is the most direct application of quantum computing. Since Aspuru-Guzik reported in *Science* that the ground-state energy of molecules in quantum computer was successfully simulated in 2005,[13] quantum chemistry simulation has immediately become the most active field in quantum computing. Therefore, many simulation software or packages for quantum computational chemistry have been developed. For example, IBM published Qiskit package in 2019 and upgraded Qiskit_nature in April, 2021,[14] which can simulate the ground state and excited state energy of molecules. In 2020, Google quantum computing team released OpenFermion package,[15] which can study the electronic structure of molecules or materials. Recently, they have developed FQE package by making full use of symmetry to reduce the cost of calculation.[16] In addition, there are JaqalPaq from Landahl's group,[17] DNA sequence reconstruction package[18] of Sarkar's group, etc. However, these programs cannot construct reasonable molecules visually and rely on one or more third-party libraries or chemical packages. Besides, there are some shortcomings, such as cumbersome functions, incomplete parameter options, and even not being provided.

Here, we release an open-source chemistry simulator ChemiQ running on quantum computer or quantum virtual machine, which is available from

https://github.com/OriginQ/QPanda-2/tree/master/QAlg/ChemiQ. The software only relies on the linear algebra library Eigen[19] and basal molecular integration library Libint2[20] in computational chemistry and does not use any third-party physical or chemical application package. By using the full-stack package, which is developed with the idea of modular programming in C++ language, one can visually define the molecular structure, simulate molecular ground-state energy, scan the curve of molecular potential energy by distance or angle, study chemical reaction, and finally analyze and display calculation results graphically.

The paper is arranged as follows: In Section II, we will illustrate the features of ChemiQ simulator. In Section III, we will briefly introduce the variational quantum eigensolver (VQE),[21-23] which is the core algorithm in the simulator. The encoding method from the fermionic Fock space to the Hilbert space of qubits and the parametrized circuits (or ansatz) will be described respectively in Section IV and V. Finally, a brief summary is given in Section VI.

## II. Features

The purpose of developing this software is to simply researchers' work (whether experts or newcomers) to carry out chemical simulation on real or virtual quantum computers, and this software has the following characteristics:

- **Modularization**. The program uses C++ object-oriented language as the development language, and highly integrates methods and functions in line with modularization.
- **Build geometry visually**. The program supports customers to construct a molecular structure visually (see Fig 1), especially for some quantum computing enthusiasts from physics, computing science, or other backgrounds than chemistry. Although such individuals are unfamiliar with chemical structures, it is no longer difficult for them to construct a reasonable molecular structure. For researchers with rich experience in chemistry, they can also use it to build the geometry they want. In addition, this function can directly import or export a file recording molecular structures which are compliable in mainstream quantum chemistry software (such as Gaussian,[24] Psi4,[25] PySCF[26], etc.).
- **Rich parameter options**. Many parameter interfaces are available to users in the program, and customers can perform calculation tasks flexibly according to their own needs.
- **Support restart or result reuse**. When the job ends abnormally, the program restarts the job based on the log or the last result.
- **Real-time analysis**. The computing results are displayed and updated in real-time on the graphical interface so that users can adjust the calculation tasks in time.
- **Cross-platform**. The software supports Windows, Linux, and Mac platforms.
- **Parallel computing**. Parallel computing is considered so that time-consuming tasks can be performed in multiprocessing or distributed parallel.

## III. Variational Quantum Eigensolver

Quantum chemistry is an exciting field of applying techniques of quantum mechanics to solve cutting-edge problems in chemical systems. More specifically, one needs to solve the Schrödinger equation which describes the motion of electrons moving around the nucleus, by considering the Coulomb force between them. The ultimate destination is to obtain the chrematistic quantum mechanical properties of the system, such as the electronic structure, the energy spectrum of this molecule, etc.

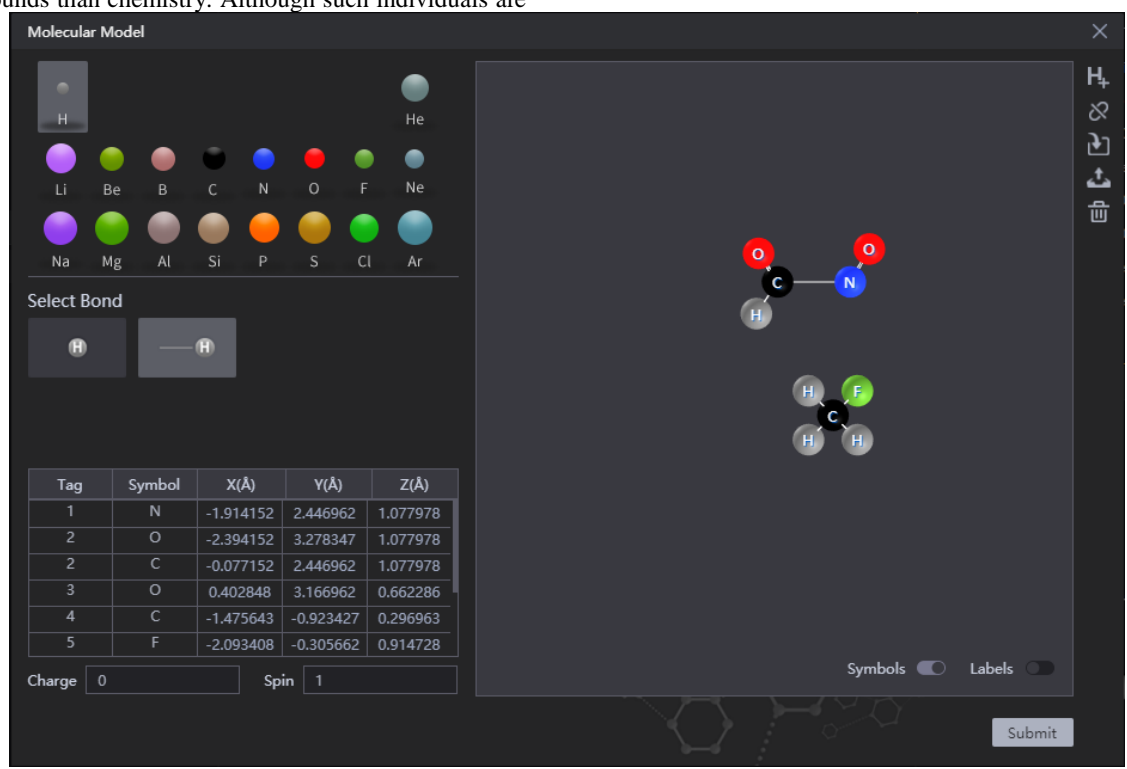

**Fig 1** The interface of building geometry visually

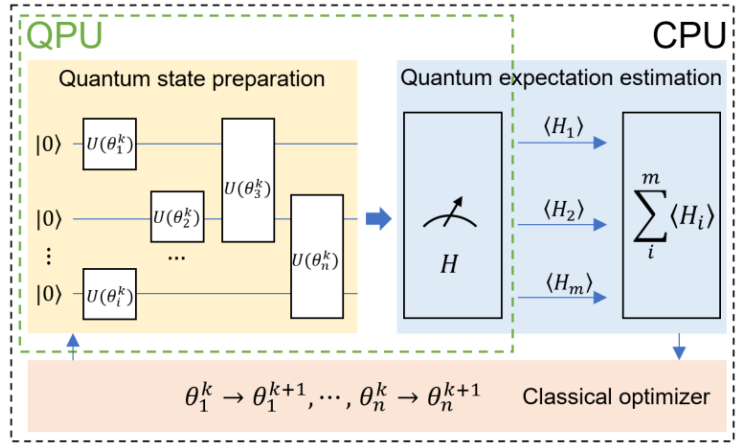

**Fig 2**. Procedure of VQE. The part in the green box is processed by a quantum computer, to prepare the parameterized wave function $\psi(\vec{\theta})\rangle$, and to measure the expectation value of the Hamiltonian $E(\vec{\theta}) = \langle\psi(\vec{\theta})|H|\psi(\vec{\theta})\rangle$, respectively. The result of measurement and the parameters $(E(\vec{\theta}), \vec{\theta})$ will be sent to a classical optimizer, the optimizer will update the parameters to minimize $E(\vec{\theta})$ and feed it back to quantum computing. After a few iterations, $E(\vec{\theta})$ will decrease to the ground state energy.

Because of the complexity of describing the wave function of the chemical system, which is exponentially proportional to size of the system, classical computers are inadequate to solve the system with good precision in an economic way. Feynman proposed that one can instead use quantum computers to simulate quantum phenomena.[2] It has been proven in some special cases quantum computers have great advantages in e.g. saving time and other resources in the calculation, such that one has "quantum advantage". As a proof-of-concept example, it is proved that the *quantum phase estimation* (QPE) algorithm can compute the energy of ground state and have an exponential acceleration compared to classical algorithms.[27] However, QPE requires too many qubits and deep quantum circuit, which is difficult to be realized in the so-called "*Noisy Intermediate-Scale Quantum*" (NISQ) era. In this era, a more practical option is to implement quantum-classical hybrid algorithm, which is usually referred to as *variational quantum eigensolver* (VQE).[22]

VQE uses a quantum computer to prepare and manipulate a parameterized wave function, and the measured result and the associated parameters are sent to a classical optimizer to determine the ideal parameters, such that the energy of the molecule is minimized:

$$E = \min_{\vec{\theta}} \langle\psi(\vec{\theta})|H|\psi(\vec{\theta})\rangle, \quad (1)$$

where $H$ is the Hamiltonian of the molecule. There are three important features of VQE: first, the execution time of quantum circuit and the number of measurements are polynomial as the size of the system increases,[28, 29] rather than exponential as in classical algorithms; second, VQE does not need quantum circuit as deep as in QPE; third, VQE is robust to quantum noise. As a result, VQE is a promising approach to realizing quantum advantage in the NISQ era.

### IV. Encoding Method

In quantum chemistry, we usually describe the systems in the second quantization formalism. The Hilbert space is spanned by $2^M$ occupation number basis states $|n_0, ..., n_{M-1}\rangle$, where $M$ is the number of fermions and $n_j \in \{0,1\}$ is the occupation number of state(orbital) $j$. Any fermionic operators can be written in terms of creation and annihilation operators $a_j^\dagger$ and $a_j$. Their action on the basis is as follows:

$$a_j^\dagger|n_0, ..., n_j, ..., n_{M-1}\rangle = \delta_{0,n_j}p_j|n_0, ..., 1_j, ..., n_{M-1}\rangle, \quad (2)$$

$$a_j|n_0, ..., n_j, ..., n_{M-1}\rangle = \delta_{1,n_j}p_j|n_0, ..., 0_j, ..., n_{M-1}\rangle, \quad (3)$$

where $p_j = (-1)^{\sum_{k=0}^{k=j-1} n_k}$ is parity factor. The creation and annihilation operators satisfy the anti-commutation relations:

$$\{a_i, a_j\} = 0, \quad \{a_i^\dagger, a_j^\dagger\} = 0, \quad \{a_i, a_j^\dagger\} = \delta_{i,j}, \quad (4)$$

where $\{A, B\}$ is anti-commutator defined by $\{A, B\} = AB + BA$. Due to the anti-commutation of fermionic systems, simulating this system on quantum computer requires mappings operators from fermionic systems to qubit systems.

Jordan-Wigner (JW) transformation[30] is a basic and widely used mapping, and Parity transformation is another basic mapping. Merging JW and Parity transformations, Bravyi-Kitaev (BK) transformation[31] improves the Pauli weight from $M$ to $\log_2 M + 1$. Note that the Pauli weight is the upper limit of number of non-trivial Pauli operators acting on the qubits. BK transformation is only suitable in case that $M$ is the power of 2, otherwise auxiliary qubits are needed to meet the condition. BK-tree transformation[32] is the generalization of BK transformation. The Pauli weight of BK-tree transformation is the same as that of BK transformation, but BK-tree transformation can be applied for any $M$ in abrupt exponential. These mappings and other mappings[32, 33] are all based on JW transformation and use no auxiliary qubits. Another class of mappings[33-36] is also based on JW transformation, but they use auxiliary qubits to eliminate a series of Pauli Z operators introduced by JW transformation. As a result, the Pauli weight is a small constant. Another transformation idea is introduced by Bravyi-Kitaev Superfast (BKSF) transformation.[31] In this mapping scheme, each state(orbital) is presented by a vertex on a graph, and the edges connecting two vertexes present the interaction terms of the Hamiltonian. Then every edge has put a qubit to simulate the fermionic systems. The Pauli weight of BKSF transformation is $d$, where $d$ is degree of the graph. However, number of qubits needed are more than $M$ in general. The BKSF transformation has generalized versions, GSEs.[37]

In ChemiQ, JW, Parity, BK, and BK-tree transformation have been realized, so we mainly introduce this class of mappings.

### IV.A JW transformation

In JW transformation, the occupation status of an orbital is stored in the qubit state, $|1\rangle$ represents occupied and $|0\rangle$ represents unoccupied. Thus the state transformation is:

$$|n_0, ..., n_j, ..., n_{M-1}\rangle \to |x_1, ..., x_j, ..., x_{M-1}\rangle, x_j = n_j \in \{0,1\}. \quad (5)$$

Let's consider the results of the creation and annihilation operators acting on the $j$th occupation state, the creation operator transforms the occupation number from 0 to 1 and eliminates 1 to nothing, and conversely, the annihilation operator transforms the occupation number from 1 to 0 and changes 0 to nothing. Here we can use a couple of single-qubit operators to present this:

$$Q_j = |0\rangle\langle 1| = \frac{X_j + iY_j}{2}, \quad Q_j^\dagger = |1\rangle\langle 0| = \frac{X_j - iY_j}{2}, \quad (6)$$

where $X_j$ and $Y_j$ are Pauli X and Y operators acting on the $j$th qubit. Besides, the creation and annihilation operators also introduce the phase factor $p_j$. The phase factor can be obtained by a string of Pauli Z operators, $Z_0 Z_1 ... Z_{j-1}$, where $Z_k$ is the Pauli Z operator acting on the $k$th qubit. The creation and annihilation operators in JW transformation are presented as:

$$a_j \to Z_0 Z_1 ... Z_{j-1} \left(\frac{X_j + iY_j}{2}\right), \quad (7)$$

$$a_j^\dagger \to Z_0 Z_1 ... Z_{j-1} \left(\frac{X_j - iY_j}{2}\right). \quad (8)$$

### IV.B Parity transformation

The parity of the $j$th state is defined as the total number of particles accumulated in the previous states. The definition of parity is the same with the phase factor $p_j$ in nature, hence we can also denote the parity of the $j$th state as $p_j = \sum_{k=0}^{k=j-1} n_k$ (mod 2). The strategy of Parity transformation is to store the total number of particles on and before the $j$th orbital by the $j$th qubit, that is $p_j + n_j$ (mod 2). The state transformation is

$$|n_0, ..., n_j, ..., n_{M-1}\rangle \to |x_1, ..., x_j, ..., x_{M-1}\rangle,$$

$$x_j = p_j + n_j = \sum_{k=0}^{k=j} n_k \text{ (mod 2)} \in \{0,1\}. \quad (9)$$

Being more accessible than the required parameters in JW transformation, the phase factor $p_j$ is easily obtained by $Z_{j-1}$. However, the representation of the action on the occupation number is more complex. Note that $x_j = p_j + n_j$, so $x_j$ is the same with $n_j$ if the parity is even. In this case, we can also use $Q_j$ and $Q_j^\dagger$ to present this. But if the parity is odd, $x_j$ is opposite to $n_j$, so the representation should exchange $Q_j$ and $Q_j^\dagger$. To determine whether the parity is even or odd, we can use the parity projectors to project the state to the even(odd) parity subspace:

$$P_{j,\text{odd}} = \left(\frac{I - Z_{j-1}}{2}\right), \quad P_{j,\text{even}} = \left(\frac{I + Z_{j-1}}{2}\right). \quad (10)$$

Then the $Q_j$ and $Q_j^\dagger$ are replaced by:

$$Q_j \to P_{j,\text{odd}} Q_j^\dagger + P_{j,\text{even}} Q_j, \quad Q_j^\dagger \to P_{j,\text{odd}} Q_j + P_{j,\text{even}} Q_j^\dagger. \quad (11)$$

Moreover, we should apply Pauli X operators on the qubits which store the particle number including the particle number of the $j$th state, so the superposition state will flip with the changing of occupation number $n_j$. All the qubits with an index greater than j store the particle number including the particle number of the $j$th state, thus we should apply Pauli X on all the qubits behind the $j$th qubit. The transformation of creation and annihilation operators is

$$a_j \to Z_{j-1}(P_{j,\text{odd}} Q_j^\dagger + P_{j,\text{even}} Q_j) X_{j+1} ... X_{M-1}$$

$$= \left(\frac{Z_{j-1} X_j + iY_j}{2}\right) X_{j+1} ... X_{M-1}, \quad (12)$$

$$a_j^\dagger \to Z_{j-1}(P_{j,\text{odd}} Q_j + P_{j,\text{even}} Q_j^\dagger) X_{j+1} ... X_{M-1}$$

$$= \left(\frac{Z_{j-1} X_j - iY_j}{2}\right) X_{j+1} ... X_{M-1}. \quad (13)$$

Although Parity transformation eliminates the Pauli Z strings, it introduces the Pauli X strings. The lengths of Pauli Z strings and Pauli X string are both $M$ at most, thus Parity transformation doesn't improve the Pauli weight compared with JW transformation.

### IV.C BK and BK-tree transformations

Improving the idea of Parity transformation, BK transformation reduces the lengths of Pauli Z and Pauli X strings to $\log_2 M$, which also reduces the Pauli weight to $\log_2 M + 1$. Suppose $M$ is in power of 2: $M = 2^d$, then the numbers 0,1,...,$M$-1 can be written in binary: $j = j_0 j_1 ... j_{d-1}$. The state transformation is

$$|n_0, ..., n_j, ..., n_{M-1}\rangle \to |x_1, ..., x_j, ..., x_{M-1}\rangle,$$

$$x_j = (n_j + \sum_{k \in S(j)} n_k)(\text{mod } 2) \in \{0,1\}, \quad (14)$$

where $S(j)$ is the summation set defined as:

$$S(j) = \{k | k \neq j \text{ and } k_l = j_l \text{ for } l \leq l_0$$

$$\text{while } j_{l_0+1} = \cdots = j_{d-1} = 1 (\text{ for some } l_0)\}. \quad (15)$$

We can write $p_j$ in terms of $x_k$: $p_j = \sum_{k \in P(k)} x_k$, where $P(j)$ is the parity set defined as:

$$P(j) = \{k | j_{l_0} = 1, k_l = j_l \text{ for } l < l_0, k_{l_0} = 0,$$

$$k_{l_0+1} = \cdots = k_{d-1} = 1 (\text{ for some } l_0)\}. \quad (16)$$

The phase factor can be obtained by $Z_{P(j)}$, where $Z_{P(j)}$ is a multi-qubit Pauli Z operator acting on the qubits belonging to the $P(j)$ set. $x_j$ can also be written in terms of $n_j$ and $x_k$: $x_j = (n_j + \sum_{k \in F(j)} x_k)(\text{mod } 2) \in \{0,1\}$, where $F(j)$ is the flip set defined as:

$$F(j) = \{k | j_{l_0+1} = \cdots = j_{d-1} = 1,$$

$$k_l = j_l, \text{for } l \neq l_0, k_{l_0} = 0 \text{ (for some } l_0)\}. \tag{17}$$

Similar to parity transformation, $Q_j$ and $Q_j^\dagger$ should be replaced by:

$$Q_j \to (\frac{I-Z_{F(j)}}{2})Q_j^\dagger + (\frac{I+Z_{F(j)}}{2})Q_j,$$

$$Q_j^\dagger \to (\frac{I-Z_{F(j)}}{2})Q_j + (\frac{I+Z_{F(j)}}{2})Q_j^\dagger. \tag{18}$$

Pauli X should be applied on the qubits which store the particle number including the particle on the $j$th orbital, and then this can be also presented as a multi-qubit Pauli operator $X_U(j)$. The $X_U(j)$ means applying Pauli X on the qubits in the $U(j)$ set, which is defined as:

$$U(j) = \{k | k \neq j \text{ and } k_l = j_l \text{ for } l \leq l_0$$

$$\text{while } k_{l_0+1} = \cdots = k_{d-1} = 1 (\text{ for some } l_0)\}. \tag{19}$$

The transformation of the creation and annihilation operators can be presented by these sets,

$$a_j \to \frac{Z_{P(j)}X_j + iZ_{P(j)-F(j)}}{2} X_{U(j)},$$

$$a_j^\dagger \to \frac{Z_{P(j)}X_j - iZ_{P(j)-F(j)}}{2} X_{U(j)}. \tag{20}$$

where $P(j) - F(j)$ is a set whose elements are in $P(j)$ but not in $F(j)$. One can also build these sets corresponding to the $\beta_M$ and $\pi_M$ matrix in Ref. [38].

To construct this mapping, we have supposed that $M$ is in power of 2, while BK-tree transformation can overcome this condition. Researchers introduced Fenwick algorithm to build the Fenwick tree, so that these sets can be easily built via the Fenwick tree. We give an example of the Fenwick tree of 8-fermion.

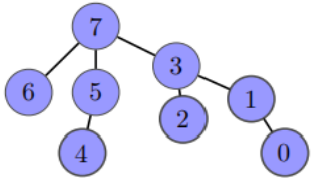

**Fig. 3** The Fenwch tree of 8-fermion.

The elements of the $F(j)$ is corresponding to the child nodes of node $j$, and the elements of $S(j)$ is the descent node of node $j$. The elements of $U(j)$ is the ancestor node of node $j$, while the elements of $P(j)$ isterminal node with index greater than $j$.

Ref. [39] introduced the MSP transformation which can be seen as the generalization of BK-tree transformation. MSP transformation can be reduced to JW, BK and other transformations with proper parameters. The authors conclude the representation of these transformations in a unified form as Eq. (20). The differences of these mappings are the differences of these sets. They give an algorithm to build these sets.

## V. Ansatz

In variational quantum algorithms (VQA), after the Hamiltonian $H$ is obtained, the selection of ansatz is an important work. Generally, the ansatz is generated with a parameterized quantum circuit (PQC) acting on an initial state $|\psi_0\rangle$:

$$|\psi(\theta)\rangle = U(\theta)|\psi_0\rangle$$

where $\theta \equiv \{\theta_1, \cdots, \theta_m\}$ is a real-valued vector of parameters. There are various ansatzes to select and are generally defined as two classes: the problem-inspired ansatz and the agnostic ansatz. The problem-inspired ansatz can consider the known information about the problem. While the agnostic ansatz generally has strong expressive power when there is little information.

### V.A Unitary coupled-cluster (UCC) ansatz

Since the coupled cluster method[40] and its variants[41-45] have achieved great success in describing chemical systems, the UCC method[46-48] has become the most well-known ansatz in quantum chemistry simulations. Given the molecular system under Hartree-Fock (HF) approximations, The HF state is the approximation of the ground state where all the electrons are lying in the orbitals with the lowest energy. When a more accurate ground state is desired, one can excite the electrons in the HF state to higher HF-energy-orbitals. The UCC ansatz has the form,

$$|\psi(\theta)\rangle = e^{T-T^\dagger}|\psi_{HF}\rangle,$$

where $|\psi_{HF}\rangle$ is the HF state and $T = \sum_k T_k$ is the cluster operator with $T_k$ the excitation operators. Considering all the excitations is resource-expensive. The UCC singles and doubles (UCCSD) are usually used,

$$T = T_1 + T_2,$$

$$T_1 = \sum_{ia} \theta_{ia} a_a^\dagger a_i, \quad T_2 = \sum_{ijab} \theta_{ijab} a_a^\dagger a_b^\dagger a_j a_i,$$

where $a_i^\dagger (a_i)$ are creation (annihilation) operators.

After the fermionic-to-qubit mapping and the *Trotter* expansion is applied, we can obtain the ansatz as

$$U(\theta) = \prod_j e^{-i\theta_j P_j},$$

Then the ansatz can be sequentially performed. This ansatz preserves the symmetry of the problem. However, the circuit always has a deep depth.

### V.B Symmetry-preserved (SP) ansatz

This is a particular problem-inspired ansatz that is suitable when the basis in the Hilbert space for the system has the same Hamming weight. That is, in the binary representation: $|i_0 i_1 \cdots i_{n-1}\rangle, i_k = \{0,1\}$, we have $\sum_k i_k = m$. The molecular system with fixed particle numbers under the Jordan-Wigner transformation exactly follows this property.

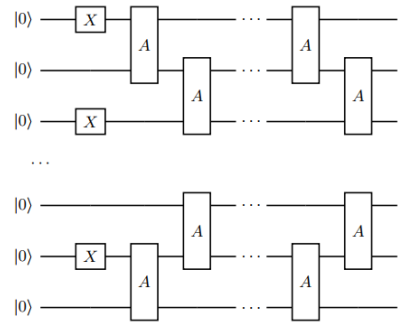

**Fig 4**. The symmetry-preserved ansatz

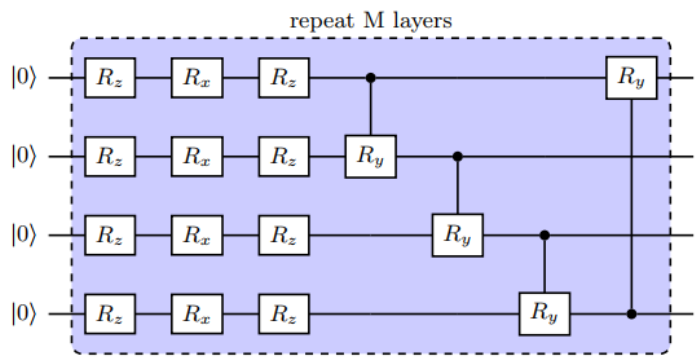

**Fig. 5** The hardware-efficient ansatz

The SP ansatz[49] takes advantage of the following 2-qubit block as components,

$$A(\theta, \phi) = \begin{pmatrix} 1 & 0 & 0 & 0 \\ 0 & \cos\theta & e^{i\phi}\sin\theta & 0 \\ 0 & e^{-i\phi}\sin\theta & -\cos\theta & 0 \\ 0 & 0 & 0 & 1 \end{pmatrix}.$$

It is obvious that block A acts identically on $|00\rangle$ and $|11\rangle$. And it mixes the state $|01\rangle$ and $|10\rangle$. This preserves symmetry. Since for n orbitals and m particles, the total freedom is $2C_n^m - 2$.. We can first set inputs as $|0\rangle^{\otimes n}$, and apply m X gates to create the producing state. Then add enough blocks in alternating layers until total freedom is achieved.

### V.C Hardware-efficient ansatz

This is a frequently used ansatz in various VQAs. When we know little information about the problem, we can use the elementary quantum gates to search for the target. Generally, the hardware-efficient ansatz[50] consists of many layers, each of which is composed of single-qubit rotations (like $R_Z R_X R_Z$) on every qubit and two-qubit entangling gates (like CZ and CNOT) on neighboring qubits. It has shown that this ansatz has strong performance, which can give us confidence that the target is in the representing area of the ansatz.

However, as shown recently, hardware-efficient ansatz with random initial parameters will suffer from the barren plateaus, where the gradient of the cost function vanishes exponentially with the number of qubits, making it difficult to train the parameters.

### VI. Conclusion

In this paper, we release the chemistry simulation software ChemiQ supported quantum computer or quantum virtual machine. It is an open-source full-stack C++ project, which only relies on the linear algebra library Eigen and the molecular integration library Libint2 of computational chemistry, and does not use any third-party application library. The software can visually construct the molecular structure, quickly simulate the molecular ground state energy, scan the molecular potential energy curve by distance or angle, study the chemical reaction, and finally analyze and display the calculation results graphically.

The current version has only been tested on the quantum virtual machine. Later, we will consider the quantum noise and access the real quantum computer.

**AUTHOR INFORMATION**

**Corresponding Authors**

*Email: gpguo@ustc.edu.cn

**ORCID**

Qingchun Wang: 0000-0003-0966-9782
Huan-Yu Liu: 0000-0002-6158-9627
Qing-Song Li: 0000-0002-8133-4939
Guo-Ping Guo: 0000-0002-2179-9507

**ACKNOWLEDGMENT**

This work was supported by the National Key Research and Development Program of China (Grant No.2016YFA0301700), the National Natural Science Foundation of China (Grants No. 61674132, 11674300, 11625419 and 11574356), the Strategic Priority Research Program of the CAS (Grant Nos. XDB24030601 and XDB30000000), the Anhui initiative in Quantum Information Technologies (Grants No. AHY080000).

**REFERENCES**

1. Moore, G. E., Cramming More Components Onto Integrated Circuits. *Electronics* **1965,** *38*, 114.
2. Feynman, R. P., Simulating physics with computers. *Int. J. Theor. Phys.* **1982,** *21* (6-7), 467-488.
3. Deutsch, D.; Jozsa, R., Rapid solution of problems by quantum computation. *Proc. R. Soc. London Ser. A: Math. Phys. Sci.* **1992,** *439* (1907), 553-558.
4. Grover, L. K. In *A fast quantum mechanical algorithm for database search*, Proceedings of the twenty-eighth annual ACM symposium on Theory of computing, 1996; pp 212-219.
5. Shor, P. W., Polynomial-Time Algorithms for Prime Factorization and Discrete Logarithms on a Quantum Computer. *SIAM J. Comput.* **1997,** *26* (5), 1484-1509.
6. Arute, F.; Arya, K.; Babbush, R., et al., Quantum supremacy using a programmable superconducting processor. *Nature* **2019,** *574* (7779), 505-510.
7. Lloyd, S., Universal Quantum Simulators. *Science* **1996,** *273* (5278), 1073-8.
8. Childs, A. M.; van Dam, W., Quantum algorithms for algebraic problems. *Rev. Mod. Phys.* **2010,** *82* (1), 1-52.
9. Ladd, T. D.; Jelezko, F.; Laflamme, R., et al., Quantum computers. *Nature* **2010,** *464* (7285), 45-53.
10. Fingerhuth, M.; Babej, T.; Wittek, P., Open source software in quantum computing. *PLoS One* **2018,** *13* (12), e0208561.
11. Abhijith, J.; Adedoyin, A.; Ambrosiano, J., et al. Quantum Algorithm Implementations for Beginners 2018, p. arXiv:1804.03719.

https://ui.adsabs.harvard.edu/abs/2018arXiv180403719A (accessed April 01, 2018).

12. Preskill, J., Quantum Computing in the NISQ era and beyond. *Quantum* **2018,** *2*, 79.

13. Aspuru-Guzik, A.; Dutoi, A. D.; Love, P. J., et al., Simulated quantum computation of molecular energies. *Science* **2005,** *309* (5741), 1704-7.

14. Aleksandrowicz, G.; Alexander, T.; Barkoutsos, P., et al., Qiskit: An open-source framework for quantum computing. *Accessed on: Mar 2019,* *16*.

15. McClean, J. R.; Rubin, N. C.; Sung, K. J., et al., OpenFermion: the electronic structure package for quantum computers. *Quantum Sci. Technol.* **2020,** *5* (3), 034014.

16. Rubin, N. C.; Shiozaki, T.; Throssell, K., et al. The Fermionic Quantum Emulator 2021, p. arXiv:2104.13944. https://ui.adsabs.harvard.edu/abs/2021arXiv210413944R (accessed April 01, 2021).

17. Maupin, O. G.; Baczewski, A. D.; Love, P. J., et al., Variational Quantum Chemistry Programs in JaqalPaq. *Entropy (Basel)* **2021,** *23* (6).

18. Sarkar, A.; Al-Ars, Z.; Bertels, K., QuASeR: Quantum Accelerated de novo DNA sequence reconstruction. *PLoS One* **2021,** *16* (4), e0249850.

19. Guennebaud, G.; Jacob, B. *Eigen v3*, http://eigen.tuxfamily.org: 2010.

20. Valeev, E. F. *Libint: high-performance library for computing Gaussian integrals in quantum mechanics*, https://github.com/evaleev/libint: 2004.

21. Yung, M. H.; Casanova, J.; Mezzacapo, A., et al., From transistor to trapped-ion computers for quantum chemistry. *Sci. Rep.* **2014,** *4* (1).

22. Peruzzo, A.; McClean, J.; Shadbolt, P., et al., A variational eigenvalue solver on a photonic quantum processor. *Nat. Commun.* **2014,** *5*, 4213.

23. McClean, J. R.; Romero, J.; Babbush, R., et al., The theory of variational hybrid quantum-classical algorithms. *New J. Phys.* **2016,** *18* (2).

24. Frisch, M. J.; Trucks, G. W.; Schlegel, H. B., et al. *Gaussian 16 Rev. A.03*, Wallingford, CT, 2016.

25. Turney, J. M.; Simmonett, A. C.; Parrish, R. M., et al., Psi4: an open-source ab initio electronic structure program. *WIREs Comput. Mol. Sci.* **2012,** *2* (4), 556-565.

26. Sun, Q.; Berkelbach, T. C.; Blunt, N. S., et al., PySCF: the Python-based simulations of chemistry framework. *WIREs Comput. Mol. Sci.* **2018,** *8* (1), e1340.

27. Abrams, D. S.; Lloyd, S., Quantum Algorithm Providing Exponential Speed Increase for Finding Eigenvalues and Eigenvectors. *Phys. Rev. Lett.* **1999,** *83* (24), 5162-5165.

28. O'Gorman, B.; Huggins, W. J.; Rieffel, E. G., et al. Generalized swap networks for near-term quantum computing 2019, p. arXiv:1905.05118. https://ui.adsabs.harvard.edu/abs/2019arXiv190505118O (accessed May 01, 2019).

29. Motta, M.; Ye, E.; McClean, J. R., et al., Low rank representations for quantum simulation of electronic structure. *npj Quantum Inf.* **2021,** *7* (1), 83.

30. Jordan, P.; Wigner, E., ber das Paulische quivalenzverbot. *Zeitschrift fr Physik* **1928,** *47* (9-10), 631-651.

31. Bravyi, S. B.; Kitaev, A. Y., Fermionic Quantum Computation. *Ann. Phys.* **2002,** *298* (1), 210-226.

32. Havlíček, V.; Troyer, M.; Whitfield, J. D., Operator locality in the quantum simulation of fermionic models. *Phys. Rev. A* **2017,** *95* (3), 032332.

33. Steudtner, M.; Wehner, S., Quantum codes for quantum simulation of fermions on a square lattice of qubits. *Phys. Rev. A* **2019,** *99* (2), 022308.

34. Verstraete, F.; Cirac, J. I., Mapping local Hamiltonians of fermions to local Hamiltonians of spins. *J. Stat. Mech.: Theory Exp.* **2005,** *2005* (09), P09012-P09012.

35. Whitfield, J. D.; Havlíček, V.; Troyer, M., Local spin operators for fermion simulations. *Phys. Rev. A* **2016,** *94* (3).

36. Farrelly, T. C.; Short, A. J., Causal fermions in discrete space-time. *Phys. Rev. A* **2014,** *89* (1), 012302.

37. Setia, K.; Bravyi, S.; Mezzacapo, A., et al., Superfast encodings for fermionic quantum simulation. *Phys. Rev. Res.* **2019,** *1* (3), 033033.

38. Seeley, J. T.; Richard, M. J.; Love, P. J., The Bravyi-Kitaev transformation for quantum computation of electronic structure. *J. Chem. Phys.* **2012,** *137* (22), 224109.

39. Li, Q.-S.; Liu, H.-Y.; Wang, Q., et al., Multilayer Segmented Parity transformation mapping fermion to qubit. 2021; p In preparation.

40. Bartlett, R. J.; Musiał, M., Coupled-cluster theory in quantum chemistry. *Rev. Mod. Phys.* **2007,** *79* (1), 291-352.

41. Offermann, R., Degenerate many fermion theory in expS form. *Nucl. Phys. A* **1976,** *273* (2), 368-382.

42. Soliverez, C. E., General theory of effective Hamiltonians. *Phys. Rev. A* **1981,** *24* (1), 4-9.

43. Hubac, I. I.; Neogrady, P., Size-consistent Brillouin-Wigner perturbation theory with an exponentially parametrized wave function: Brillouin-Wigner coupled-cluster theory. *Phys. Rev. A* **1994,** *50* (6), 4558-4564.

44. Mahapatra, U. S.; Datta, B.; Mukherjee, D., A size-consistent state-specific multireference coupled cluster theory: Formal developments and molecular applications. *J. Chem. Phys.* **1999,** *110* (13), 6171-6188.

45. Wang, Q.; Duan, M.; Xu, E., et al., Describing Strong Correlation with Block-Correlated Coupled Cluster Theory. *J. Phys. Chem. Lett.* **2020,** *11* (18), 7536-7543.

46. Yaris, R., Linked‐Cluster Theorem and Unitarity. *J. Chem. Phys.* **1964,** *41* (8), 2419-2421.

47. Yaris, R., Cluster Expansions and the Unitary Group. *J. Chem. Phys.* **1965,** *42* (9), 3019-3024.

48. Taube, A. G.; Bartlett, R. J., New perspectives on unitary coupled-cluster theory. *Int. J. Quantum Chem.* **2006,** *106* (15), 3393-3401.

49. Gard, B. T.; Zhu, L.; Barron, G. S., et al., Efficient symmetry-preserving state preparation circuits for the variational quantum eigensolver algorithm. *npj Quantum Inf.* **2020,** *6* (1).

50. Kandala, A.; Mezzacapo, A.; Temme, K., et al., Hardware-efficient variational quantum eigensolver for small molecules and quantum magnets. *Nature* **2017,** *549* (7671), 242-246.